\documentclass[12pt]{article}
\def\be{\begin{equation}}
\def\ee{\end{equation}}
\def\bea{\begin{eqnarray}}
\def\eea{\end{eqnarray}}
\usepackage{graphicx}

\catcode`\@=11
\def\lsim{\mathrel{\mathpalette\@versim<}}
\def\gsim{\mathrel{\mathpalette\@versim>}}
\def\@versim#1#2{\vcenter{\offinterlineskip
\ialign{$\m@th#1\hfil##\hfil$\crcr#2\crcr\sim\crcr } }}
\catcode`\@=12
\input epsf.tex

\catcode`@=12
\begin{document}
\thispagestyle{empty}
\begin{flushright}
UCRHEP-T501\\
January 2011\
\end{flushright}
\vspace{0.3in}
\begin{center}
{\LARGE \bf Observable Lepton Flavor Symmetry at LHC\\}
\vspace{1.2in}
{\bf Ernest Ma\\}
\vspace{0.2in}
{\sl Department of Physics and Astronomy, University of California,\\ 
Riverside, California 92521, USA\\}
\end{center}
\vspace{1.0in}
\begin{abstract}\
I discuss a model of lepton flavor symmetry based on the non-Abelian finite 
group $T_7$ and the gauging of $B-L$, which has a residual $Z_3$ symmetry 
in the charged-lepton Yukawa sector, allowing it to be observable at the 
Large Hadron Collider (LHC) from the decay of the new $Z'$ gauge boson 
of this model to a pair of scalar bosons which have the unusual highly 
distinguishable final states $\tau^- \tau^- \mu^+ e^+$.
\end{abstract}
\vspace{1.0in}
\noindent Talk at the ``International Conference on Flavor Physics in the 
LHC Era,'' Singapore (November 2010).

\newpage
\baselineskip 24pt

\section{A Short History of $A_4$}

In 1978, soon after the putative discovery of the third family of leptons 
and quarks, it was conjectured by Cabibbo\cite{c78} and Wolfenstein\cite{w78} 
independently that the $3 \times 3$ lepton mixing matrix may be given by
\begin{equation}
U^{CW}_{l \nu} = {1 \over \sqrt{3}} \pmatrix{1 & 1 & 1 \cr 1 & \omega & \omega^2 
\cr 1 & \omega^2 & \omega},
\end{equation}
where $\omega = \exp(2\pi i/3) = -1/2 + i \sqrt{3}/2$.  This implies 
$\theta_{12} = \theta_{23} = \pi/4$, $\tan^2 \theta_{13} = 1/2$, and 
$\delta_{CP} = \pm \pi/2$.  Thirty years later, we know that they were not 
completely correct, but their bold conjecture illustrated the important 
point that not everyone expected small mixing angles in the lepton sector 
as in the quark sector.  The fact that neutrino mixing turns out to involve 
large angles should not have been such a big surprise.

In 2001, Ma and Rajsekaran\cite{mr01} showed that the non-Abelian discrete 
symmetry $A_4$ allows $m_{e,\mu,\tau}$ to be arbitrary, and yet 
$\sin^2 2 \theta_{atm} = 1$, $\theta_{e3} = 0$ can be obtained.  In 2002, 
Babu, Ma, and Valle\cite{bmv03} showed how $\theta_{13} \neq 0$ can be 
radiatively generated in $A_4$ with the prediction that $\delta_{CP} = \pm 
\pi/2$, i.e. maximum $CP$ violation.

In 2002, Harrison, Perkins, and Scott\cite{hps02}, after abandoning their 
bimaximal and trimaximal hypotheses, proposed the tribimaximal mixing 
matrix, i.e.
\begin{equation}
U^{HPS}_{l \nu} = \pmatrix{\sqrt{2/3} & 1/\sqrt{3} & 0 \cr -1/\sqrt{6} & 
1/\sqrt{3} & -1/\sqrt{2} \cr -1/\sqrt{6} & 1/\sqrt{3} & 1/\sqrt{2}} \sim 
(\eta_8, \eta_1,\pi^0),
\end{equation}
which is easy to remember in terms of the meson nonet.  This means that 
$\sin^2 2 \theta_{atm} = 1$, $\tan^2 \theta_{sol} = 1/2$, $\theta_{e3} = 0$.

In 2004, I showed\cite{m04} that tribimaximal mixing may be obtained in 
$A_4$, with
\begin{equation}
U^\dagger_{CW} M_\nu U_{CW} = \pmatrix{a+2b & 0 & 0 \cr 0 & a-b & d \cr 
0 & d & a-b},
\end{equation}
in the basis that $M_l$ is diagonal.  At that time, the Sudbury Neutrino 
Observatory (SNO) data gave $\tan^2 \theta_{sol} = 0.40 \pm 0.05$, but it 
was changed in early 2005 to $0.45 \pm 0.05$.  Thus tribimaximal mixing 
and $A_4$ became part of the lexicon of the neutrino theorist.

After the 2005 SNO revision, two $A_4$ models quickly appeared. (I) 
Altarelli and Feruglio\cite{af05} proposed
\begin{equation}
U^\dagger_{CW} M_\nu U_{CW} = \pmatrix{a & 0 & 0 \cr 0 & a & d \cr 
0 & d & a},
\end{equation}
i.e. $b=0$, and (II) Babu and He\cite{bh05} proposed
\begin{equation}
U^\dagger_{CW} M_\nu U_{CW} = \pmatrix{a'-d^2/a' & 0 & 0 \cr 0 & a' & d \cr 
0 & d & a'},
\end{equation}
i.e. $d^2 = 3b(b-a)$.

The {\it challenge} is to prove experimentally that $A_4$ or some other 
discrete symmetry is behind neutrino tribimaximal mixing.  If $A_4$ is 
realized by a renormalizable theory at the electroweak scale, then the 
extra Higgs doublets required will bear this information.  Specifically, 
$A_4$ breaks to the residual symmetry $Z_3$ in the charged-lepton sector, 
and all Higgs Yukawa interactions are determined in terms of lepton masses. 
This notion of {\it lepton flavor triality}\cite{m10} may be the key 
to such a proof, but these exotic Higgs doublets are very hard to see at 
the LHC.

\section{Frobenius Group $T_7$}

The tetrahedral group $A_4$ (12 elements) is the smallest group with a real 
\underline{3} representation. The Frobenius group $T_7$ (21 elements) is 
the smallest group with a pair of complex \underline{3} and \underline{3}$^*$ 
representations.  It is generated by
\begin{equation}
a = \pmatrix{\rho & 0 & 0 \cr 0 & \rho^2 & 0 \cr 0 & 0 & \rho^4}, ~~~ 
b = \pmatrix{0 & 1 & 0 \cr 0 & 0 & 1 \cr 1 & 0 & 0},
\end{equation}
where $\rho = \exp(2 \pi i /7)$, so that $a^7=1$, $b^3=1$, and $ab = ba^4$. 
It has been considered by Luhn, Nasri, and Ramond\cite{lnr07}, Hagedorn, 
Schmidt, and Smirnov\cite{hss09}, as well as King and Luhn\cite{kl09}. 
The character table of $T_7$ (with $\xi = -1/2 + i \sqrt{7}/2$) is given by

\begin{table}[htb]
\centerline{\begin{tabular}{|c|c|c|c|c|c|c|c|}
\hline
class & $n$ & $h$ & $\chi_1$ & $\chi_{1'}$ & $\chi_{1''}$ & $\chi_3$ & 
$\chi_{3^*}$ \\
\hline 
$C_1$ & 1 & 1 & 1 & 1 & 1 & 3 & 3 \\
$C_2$ & 7 & 3 & 1 & $\omega$ & $\omega^2$ & 0 & 0 \\
$C_3$ & 7 & 3 & 1 & $\omega^2$ & $\omega$ & 0 & 0 \\
$C_4$ & 3 & 7 & 1 & 1 & 1 & $\xi$ & $\xi^*$ \\
$C_5$ & 3 & 7 & 1 & 1 & 1 & $\xi^*$ & $\xi$ \\
\hline
\end{tabular}}
\caption{Character table of $T_7$.}
\end{table}

The group multiplication rules of $T_7$ include
\begin{eqnarray}
\underline{3} \times \underline{3} &=& \underline{3}^* (23,31,12) + 
\underline{3}^* (32,13,21) + \underline{3} (33,11,22), \\  
\underline{3} \times \underline{3}^* &=& \underline{3} (2 1^*, 3 2^*, 1 3^*) + 
\underline{3}^* (1 2^*, 2 3^*, 3 1^*) + \underline{1} (1 1^* + 2 2^* + 3 3^*) 
\nonumber \\  &+& \underline{1}' (1 1^* + \omega 2 2^* + \omega^2 3 3^*) +
\underline{1}'' (1 1^* + \omega^2 2 2^* + \omega 3 3^*).
\end{eqnarray}  
Note that $\underline{3} \times \underline{3} \times \underline{3}$ has two 
invariants and $\underline{3} \times \underline{3} \times \underline{3}^*$ 
has one invariant. These serve to distinguish $T_7$ from $A_4$ and $\Delta(27)$.

\section{$U(1)_{B-L}$ Gauge Extension with $T_7$}

Recently, the following model has been proposed by Cao, Khalil, Ma, and 
Okada\cite{ckmo10}:  Under $T_7$, let $L_i = (\nu,l)_i \sim \underline{3}$, 
$l^c_i \sim \underline{1},\underline{1}',\underline{1}''$,  $\Phi_i = 
(\phi^+,\phi^0)_i \sim \underline{3}$, which means that $\tilde{\Phi} = 
(\bar{\phi}^0, -\phi^-)_i \sim \underline{3}^*$.  The Yukawa couplings 
$L_i l^c_j \tilde{\Phi}_k$ generate the charged-lepton mass matrix
\begin{equation}
M_l = \pmatrix{f_1 v_1 & f_2 v_1 & f_3 v_1 \cr f_1 v_2 & \omega^2 f_2 v_2 & 
\omega f_3 v_2 \cr f_1 v_3 & \omega f_2 v_3 & \omega^2 f_3 v_3} = 
U^\dagger_{CW} \pmatrix{f_1 & 0 & 0 \cr 0 & f_2 & 0 \cr 0 & 0 & f_3} \sqrt{3}v,
\end{equation}
if $v_1=v_2=v_3=v$ as in the original $A_4$ proposal. 

Let $\nu^c_i \sim \underline{3}^*$, then the Yukawa couplings $L_i \nu^c_j 
\Phi_k$ are allowed, with
\begin{equation}
M_D = f_D v \pmatrix{0 & 1 & 0 \cr 0 & 0 & 1 \cr 1 & 0 & 0}.
\end{equation}
Note that $\Phi$ and $\tilde{\Phi}$ have $B-L=0$.

Now add the neutral Higgs singlets $\chi_i \sim \underline{3}$ and $\eta \sim 
\underline{3}^*$, both with $B-L = -2$.  Then there are two Yukawa invariants: 
$\nu^c_i \nu^c_j \chi_k$ and $\nu^c_i \nu^c_j \eta_k$.  Note that $\chi_i^* \sim 
\underline{3}^*$ is not the same as $\eta_i \sim \underline{3}^*$ because they 
have different $B-L$.  This means that both $B-L$ and the complexity of $T_7$ 
are required for this scenario.  The heavy Majorana mass matrix for $\nu^c$ 
is then
\begin{equation}
M = h \pmatrix{u_2 & 0 & 0 \cr 0 & u_3 & 0 \cr 0 & 0 & u_1} + 
h' \pmatrix{0 & u'_3 & u'_2 \cr u'_3 & 0 & u'_1 \cr u'_2 & u'_1 & 0} = 
\pmatrix{A & 0 & B \cr 0 & A & 0 \cr B & 0 & A},
\end{equation}
where $A=hu_1=hu_2=hu_3$ and $B=h'u'_2$ with $u'_1=u'_3=0$ have been assumed, 
i.e. $\chi_i$ breaks in the (1,1,1) direction, whereas $\eta_i$ breaks in the 
(0,1,0) direction.  This is the $Z_3-Z_2$ misalignment also used in $A_4$ 
models.  The seesaw neutrino mass matrix is now
\begin{equation}
M_\nu = -M_D M^{-1} M_D^T = {-f_D^2 v^2 \over A^3 - AB^2} \pmatrix{A^2-B^2 & 
0 & 0 \cr 0 & A^2 & -AB \cr 0 & -AB & A^2},
\end{equation}
i.e. the two-parameter tribimaximal form proposed by Babu and He, but 
without the auxiliary $Z_4 \times Z_3$ symmetry assumed there.  Two limiting 
cases are (I) normal hierarchy $(d = -a)$: $m_1=m_2=0$, $m_3=2a$, and (II) 
inverted hierarchy $(d=2a)$: $m_1=3a$, $m_2=-3a$, $m_3=-a$, with the 
effective mass in neutrinoless double beta decay given by $m_{ee} = a = 
\sqrt{\Delta m^2_{atm}/8} = 0.02$ eV.

\section{Higgs Structure}

In the charged-lepton Yukawa sector, i.e. $L_i  l^c_j \tilde{\Phi}_k$, a 
residual $Z_3$ symmetry exists so that linear combinations of $\Phi_k$ 
become $\phi_0,\phi_1,\phi_2 \sim 1, \omega, \omega^2$ together with 
$e,\mu,\tau \sim 1, \omega^2, \omega$.  Their interactions are given by
\begin{eqnarray}
{\cal L}_Y &=& (\sqrt{3}v)^{-1} [m_\tau \bar{L}_\tau \tau_R +  
m_\mu \bar{L}_\mu \mu_R + m_e \bar{L}_e e_R] \phi_0 \nonumber \\ 
&+& (\sqrt{3}v)^{-1} [m_\tau \bar{L}_\mu \tau_R +  
m_\mu \bar{L}_e \mu_R + m_e \bar{L}_\tau e_R] \phi_1 \nonumber \\ 
&+& (\sqrt{3}v)^{-1} [m_\tau \bar{L}_e \tau_R +  
m_\mu \bar{L}_\tau \mu_R + m_e \bar{L}_\mu e_R] \phi_2 + H.c.\\ 
\end{eqnarray}
As a result, the rare decays $\tau^+ \to \mu^+ \mu^+ e^-$ and $\tau^+ \to 
e^+ e^- \mu^-$ are allowed, but no others.  For example, $\mu \to e \gamma$ 
is forbidden.  Here $\phi_1^0, \bar{\phi}_2^0 \sim \omega$, mixing to form 
mass eigenstates $\psi^0_{1,2} = (\phi_1^0 \pm \bar{\phi}_2^0)/\sqrt{2}$. 
Using
\begin{equation}
{B(\tau^+ \to \mu^+ \mu^+ e^-) \over B(\tau \to \mu \nu \nu)} = {m_\tau^2 
m_\mu^2 (m_1^2 + m_2^2)^2 \over m_1^4 m_2^4} < {2.3 \times 10^{-8} \over 0.174},
\end{equation}
the bound $m_1 m_2/\sqrt{m_1^2+m_2^2} > 22~{\rm GeV}~
(174~{\rm GeV}/\sqrt{3} v)$ is obtained.  Hence the production of $\psi^0_{1,2} 
\bar{\psi}^0_{2,1}$ at the LHC with final states $\tau^- e^+ \tau^- \mu^+$ 
and $\tau^+ \mu^- \tau^+ e^-$ would be indicative of this $Z_3$ flavor 
symmetry.

\begin{figure}[htb]
\centerline{\includegraphics[width=12cm]{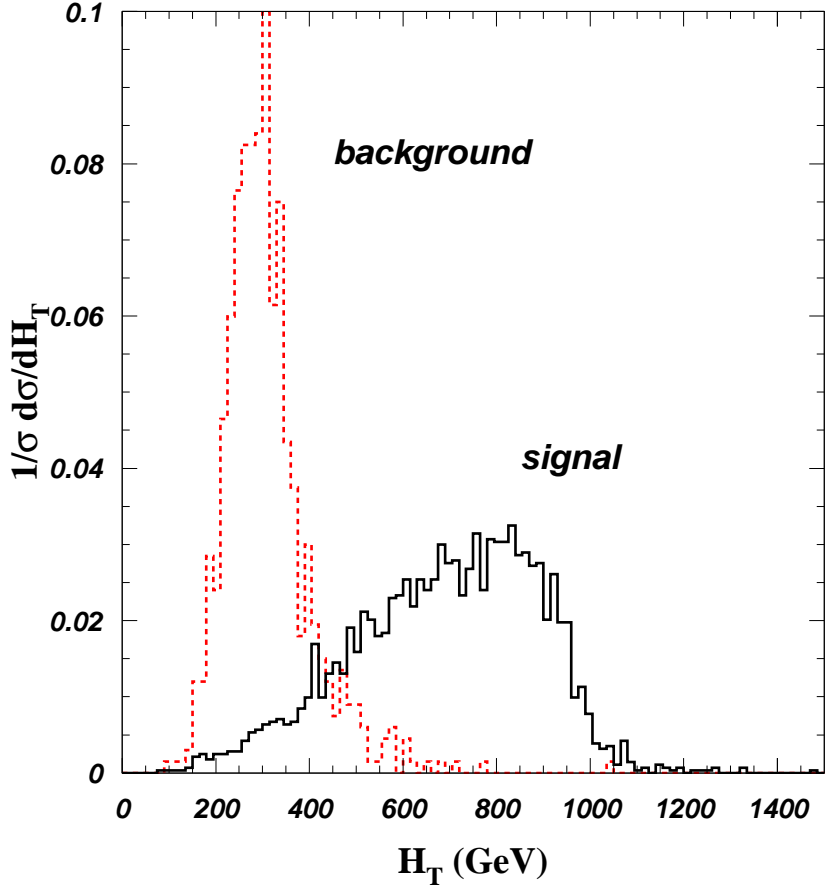}}
\vspace*{8pt}
\caption{Normalized distribution of $H_T$.}
\end{figure}

\section{LHC Observations}

The $\phi_{1,2}$ scalar doublets have $B-L=0$, so they do not couple directly 
to the $Z'_{B-L}$ gauge boson, but they can mix, after $U(1)_{B-L}$ breaking, 
with the $\chi$ and $\eta$ singlets ($B-L=-2$) which do.  Thus this model 
can be tested at the LHC by discovering $Z'$ from $q \bar{q} \to Z' \to 
\mu^- \mu^+$ and looking for $Z' \to \psi^0_{1,2} \bar{\psi}^0_{2,1}$ with 
the subsequent decays $\psi \to \tau^- e^+$ and $\bar{\psi} \to \tau^- \mu^+$.
We assume for simplicity that $m_1=m_2$, and take the LHC energy as  
14 TeV, which is expected to be reached in a year or two.

\begin{figure}[htb]
\centerline{\includegraphics[width=12cm]{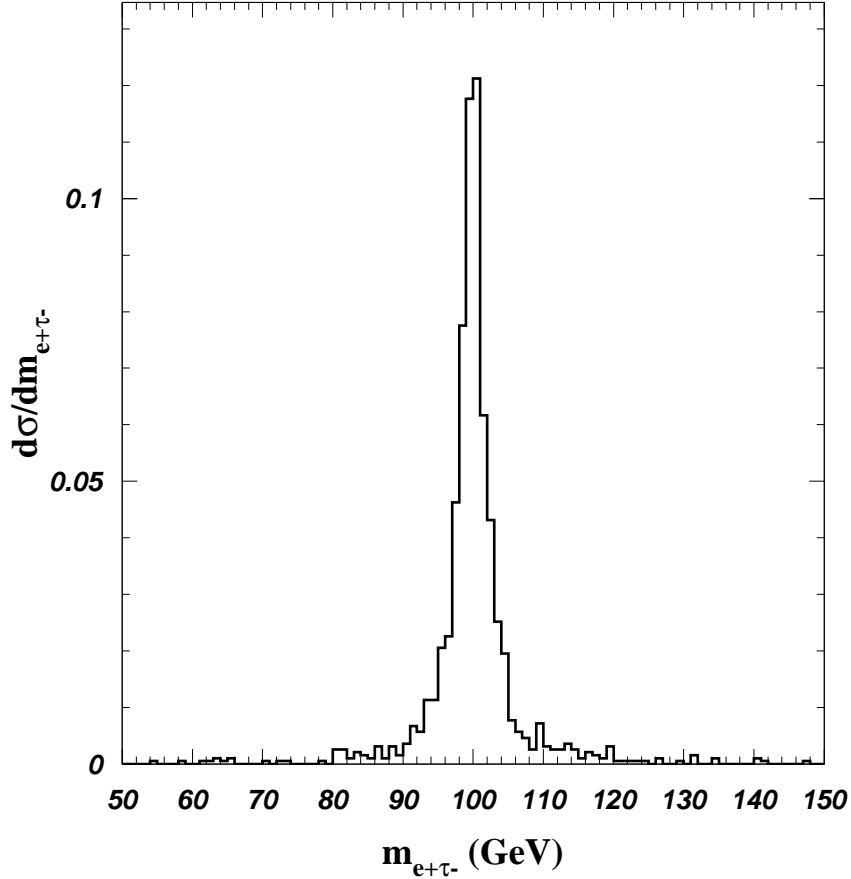}}
\vspace*{8pt}
\caption{Distribution of the invariant mass of the $e^+$ and reconstructed 
$\tau^-$ pair.}
\end{figure}

Let $\Gamma_0 = g^2_{B-L} m_{Z'}/12 \pi$, then the partial decay widths of 
$Z'$ are
\begin{eqnarray}
\Gamma_q &=& (6)(3)(1/3)^2 \Gamma_0, \\ 
\Gamma_l &=& (3) (-1)^2 \Gamma_0, \\ 
\Gamma_\nu &=& (3)(-1)^2(1/2) \Gamma_0, \\ 
\Gamma_\psi &\simeq& (2)(-2)^2 \sin^2 \theta (1/4) \Gamma_0,
\end{eqnarray}
where $\sin \theta$ is an effective parameter accounting for the mixing of 
$\psi$ to $\chi$ and $\eta$.  The signature events are chosen to be 
$\tau^- \tau^- \mu^+ e^+$ with $\tau^-$ decaying into $l^- (e^-~{\rm or}~\mu^-)$ 
plus missing energy.  The bakground events yielding this signature come from
\begin{eqnarray}
WWZ &:& pp \to W^+W^-Z, ~W^\pm \to l^\pm \nu, ~Z \to l^+l^-, \\
ZZ &:& pp \to ZZ, ~Z \to l^+l^-, ~Z \to \tau^+\tau^-, ~\tau^\pm \to 
l^\pm \nu \nu, \\ 
t \bar{t} &:& pp \to t \bar{t} \to b(\to l^-) \bar{b}(\to l^+) W^+ W^-, 
~w^\pm \to l^\pm \nu, \\ 
Z b \bar{b} &:& pp \to Z b(\to l^-) \bar{b}(\to l^+), ~Z \to l^+l^-.
\end{eqnarray}
We require no jet tagging and consider only events with both $e^+$ and 
$\mu^+$ in the final states.  Our benchmark points for $m_{Z'},m_\psi$ 
(in GeV) are (A), (1000,100), (B) (1500,100), (C) (1000,300), (D) (1500,300), 
with $g_{B-L}=g_2=e/\sin \theta_W$, and $\sin^2 \theta=0.2$.  We impose the 
following basic acceptance cuts:
\begin{eqnarray}
&& p_{T,l}^{(1,2)} > 50~{\rm GeV}, ~~~  p_{T,l}^{(3,4)} > 20~{\rm GeV}, ~~~  
|\eta_l| < 2.5, \\ 
&& \Delta R_{ij} \equiv \sqrt{(\eta_i-\eta_j)^2 + (\phi_i - \phi_j)^2} > 0.4, 
~~~ {\rm missing}~E_T > 30~{\rm GeV},
\end{eqnarray}
where $\Delta R_{ij}$ is the separation in the azimuthal angle $(\phi)~-$ 
pseudorapidity $(\eta)$ plane between $i$ and $j$.  We also model detector 
resolution effects by smearing the final-state energy.  To further suppress 
the backgrounds, we require
\begin{equation}
H_T \equiv \sum_i p_{T,i} + ~{\rm missing}~E_T > 300~{\rm GeV},
\end{equation}
where $i$ denotes the visible particles.

\begin{figure}
\centerline{\includegraphics[width=12cm]{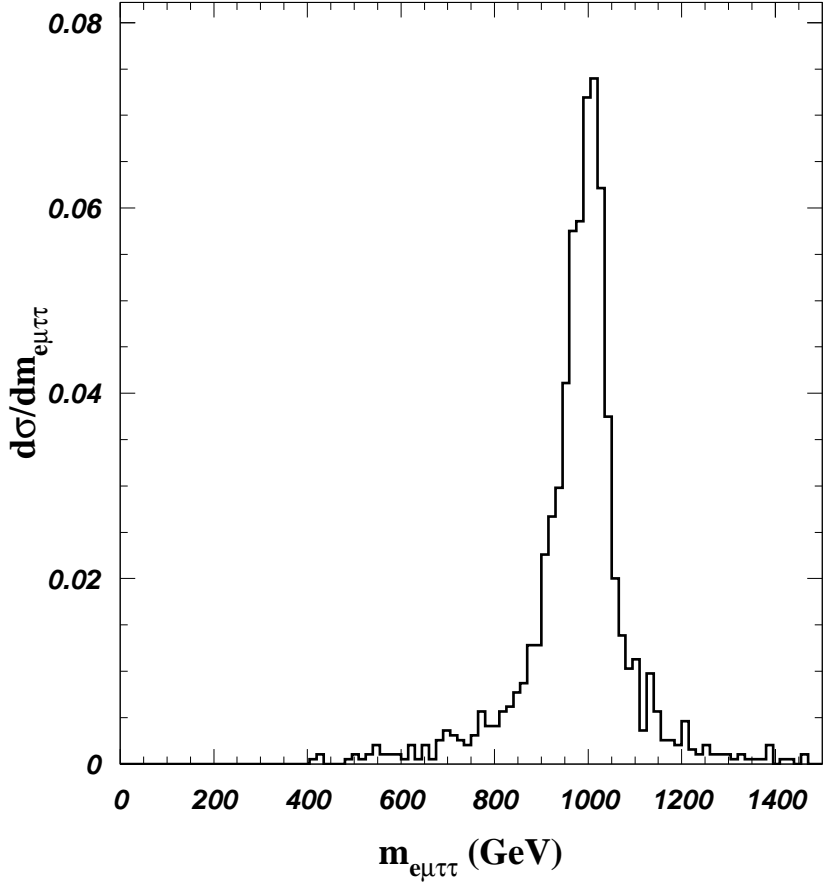}}
\vspace*{8pt}
\caption{Distribution of the reconstructed $Z'$ mass.}
\end{figure}

To reconstruct the scalar $\psi$, we adopt the collinear approximation that 
the $l$ and $\nu$'s from $\tau$ decays are parallel due to the $\tau$'s 
large boost, coming from the heavy $\psi$.  
Denoting by $x_{\tau_i}$ the 
fraction of the parent $\tau$ energy which each observable decay particle 
carries, the transverse momentum vectore are related by
\begin{equation}
{\rm missing}~E_T = (1/x_{\tau_1} - 1) \vec{p}_1 + (1/x_{\tau_2} - 1) \vec{p}_2.
\end{equation}
When the decay products are not back-to-back, this gives two conditions 
for $x_{\tau_i}$, with the $\tau$ momenta as $\vec{p}_1/x_{\tau_1}$ and 
$\vec{p}_2/x_{\tau_2}$, respectively.  We further require $x_{\tau_i} > 0$ to 
remove the unphysical solutions, and minimize $\Delta R_{e^+l^-}$ to choose 
the correct $e^+ l^-$ to reconstruct $\psi$ and then $Z'$.  In Table 2 we 
show the signal and background cross sections (in fb) for the benchmark 
cases (A) and (C).

\begin{table}[htb]
\centerline{\begin{tabular}{|c|c|c|c|c|c|c|}
\hline
 & (A) & (C) & $t \bar{t}$ & $WWZ$ & $ZZ$ & $Z b \bar{b}$ \\ 
\hline
no cut & 5.14 & 2.57 & 1.22 & 0.21 & 27.11 & 2.99 \\ 
basic cut & 1.46 & 1.05 & 0.16 & 0.02 & 0.0052 & 0.024 \\ 
$H_T$ cut & 1.41 & 1.04 & 0.08 & 0.006 & 0.0 & 0.0 \\ 
$x_\tau > 0$ & 0.69 & 0.52 & 0.015 & 0.002 & 0.0 & 0.0 \\
\hline
\end{tabular}}
\caption{Signal and background cross sections (in fb) for $m_{Z'} = 1$ TeV 
and $m_\psi = 100$ GeV (A) and 300 GeV (C).}
\end{table}

We show in Fig.~1 the $H_T$ distribution in case (A) to demonstrate the 
separation of signal from background.  We then show in Fig.~2 how the 
mass of $\psi$ may be obtained from $e^+ \tau^-$, and in Fig.~3 
how the $Z'$ mass may be reconstructed.  The 5$\sigma$ discovery 
contours for (A) to (D) are shown in Fig.~4 in the $m_{Z'}-\sin^2 \theta$ 
plane.

\begin{figure}[htb]
\centerline{\includegraphics[width=12cm]{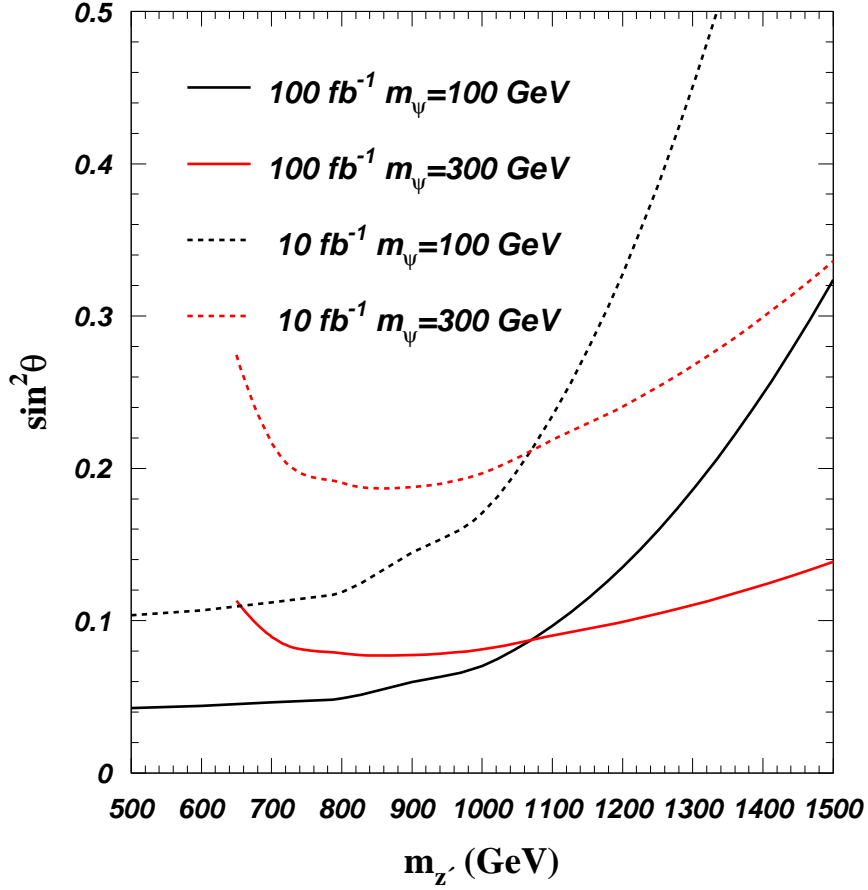}}
\vspace*{8pt}
\caption{The 5$\sigma$ discovery contours in the $m_{Z'} - \sin^2 \theta$ 
plane.}
\end{figure}

\section{Conclusion}

Using the non-Abelian discrete symmetry $T_7$ together with the gauging of 
$B-L$, a simple renormalizable two-parameter model of the neutrino mass 
matrix is obtained with tribimaximal mixing.  The charged-lepton Higgs 
Yukawa interactions are predicted completely and exhibit a residual $Z_3$ 
symmetry which is verifiable at the LHC.  The signature is $pp \to Z' \to 
\psi \bar{\psi}$ with the subsequent decays $\psi \to \tau^- e^+$ and 
$\bar{\psi} \to \tau^- \mu^+$.  With 10 fb$^{-1}$ at $E_{cm} = 14$ TeV and 
$m_{Z'} \sim 1$ TeV, a 5$\sigma$ discovery is expected.

\section*{Acknowledgements}

I thank Harald Fritzsch, K. K. Phua, and the other organizers 
for their great hospitality and a stimulating conference. This work 
is supported in part by the U.~S.~Department of Energy under Grant 
No. DE-FG03-94ER40837.


\begin{thebibliography}{0}
\bibitem{c78}
N. Cabibbo, {\it Phys. Lett.} {\bf B72}, 333 (1978).

\bibitem{w78}
L. Wolfenstein, {\it Phys. Rev.} {\bf D18}, 958 (1978).

\bibitem{mr01}
E. Ma and G. Rajasekaran, {\it Phys. Rev.} {\bf D64}, 113012 (2001).

\bibitem{bmv03}
K. S. Babu, E. Ma, and J. W. F. Valle, {\it Phys. Lett.} {\bf B552}, 
207 (2003).

\bibitem{hps02} 
P. F. Harrison, D. H. Perkins, and W. G. Scott, {\it Phys. Lett.} {\bf B530}, 
167 (2002).

\bibitem{m04} 
E. Ma, {\it Phys. Rev.} {\bf D70}, 031901 (2004).

\bibitem{af05}
G. Altarelli and F. Feruglio, {\it Nucl. Phys.} {\bf B720}, 64 (2005).

\bibitem{bh05}
K. S. Babu and X.-G. He, arXiv:0507217 [hep-ph].

\bibitem{m10}
E. Ma, {\it Phys. Rev.} {\bf D82}, 037301 (2010).

\bibitem{lnr07}
C. Luhn, S. Nasri, and P. Ramond, {\it Phys. Lett.} {\bf B652}, 27 (2007).

\bibitem{hss09}
C. Hagedorn, M. A. Schmidt, and A. Yu. Smirnov, {\it Phys. Rev.} {\bf D79}, 
036002 (2009).

\bibitem{kl09} 
S. F. King and C. Luhn, {\it JHEP} {\bf 0910}, 093 (2009).

\bibitem{ckmo10}
Q.-H. Cao, S. Khalil, E. Ma, and H. Okada, arXiv:1009.5415 [hep-ph].

\end{thebibliography}
\end{document}